# Vapor-deposited thin films with negative refractive index in the visible regime

Yi-Jun Jen,[a,c] Akhlesh Lakhtakia,[b,d] Ching-Wei Yu[a] and Chin-Te Lin[a]

[a]Department of Electro-Optical Engineering, National Taipei University of Technology, No. 1, Sec. 3, Chung-Hsiao E. Rd., Taipei 106, Taiwan

[b]Materials Research Institute and Department of Engineering Science and Mechanics, Pennsylvania State University, University Park, PA 16802, USA

[c]E-mail: jyjun@ntut.edu.tw

[d]Corresponding Author. E-mail: akhlesh@psu.edu

**ABSTRACT**

*Metamaterials are artificial composite materials that, by virtue of their microstructure, exhibit properties not exhibited by their component materials. Much excitement has been generated by negatively refracting metamaterials—typically consisting of coupled, metallic, subwavelength elements that simulate electric and magnetic dipoles—because of their unusual ability to manipulate visible light, infrared waves and microwaves.[1,2] In such a metamaterial, an electromagnetic wave propagates so that the direction of its energy flow is opposed to its phase velocity, a condition captured by the real part of the refractive index being negative. Here we show that a thin film comprising parallel tilted nanorods deposited by directing silver vapor obliquely towards a large-area substrate displays a negative real refractive index in the visible regime. Since vapor deposition is a very well-established technique to deposit thin films in the photonics industry,[3,4] our result is promising for large-scale production of negatively refracting metamaterials.*

## 1. INTRODUCTION

Negatively refracting metamaterials in the microwave and terahertz regimes are periodic arrays of subwavelength elements comprising incomplete loops and straight wires.[2,5] In the near-infrared regime (wavelength $\lambda \sim 1.5$ $\mu m$), arrays of parallel pairs of metal rods in a dielectric matrix material[6] and parallel pairs of dielectric rods in a metal matrix[7] have been used as those elements. Recently, a two-dimensionally periodic array of parallel silver nanowires embedded in alumina was shown to negatively refract in the visible regime ($\lambda = 660$ nm).[8]

Photonic applications of negative refraction shall be greatly facilitated by the availability of a simple technique that allows the fabrication of thin-film samples with large transverse area. With that goal in mind, we decided to use the oblique angle deposition (OAD) technique that emerged in the 1860s[9] and is now considered a workhorse technique in the optical-thin-film industry.[3,4]

## 2. EXPERIMENTAL PROCEDURES AND RESULTS

In an evacuated chamber, vapor from a solid is directed at an angle (deposition angle) to the normal to a flat substrate. The vapor is either thermally generated by heating the solid or by directing an energetic beam of electrons, ions, or photons towards the solid.[3,4] Initially, nucleation centers form randomly on the substrate.[10] Deposition conditions can be chosen so that, subsequently, nanorods grow preferentially towards the incoming vapor because of the self-shadowing effect.[11,12] The resulting thin film has long been recognized to be optically anisotropic,[13] like a biaxial crystal.[14]

We deposited thin films comprising parallel, tilted silver nanorods on 2-inch square substrates of fused silica by electron-beam evaporation. The chamber containing the substrate and the solid silver was pumped to a base pressure of $4 \times 10^{-6}$ Pa prior to each deposition. The deposition rate was maintained at 0.3 nm/s and the deposition angle was set at $86^\circ$. A quartz thickness monitor set next to the substrate was used to control the deposition rate and the thickness of the film.

Figure 1 presents a top-view scanning electron microscopic (SEM) image of a silver thin film deposited by us. Clearly, this thin film is an ensemble of parallel tilted nanorods. The film thickness is 240 nm, as observed from the cross-section SEM image. The angle between the normal to the substrate and the tilt of the nanorods is $66^\circ \pm 5^\circ$. The average length and diameter of the nanorods are 650 nm and 80 nm, respectively.

In order to ascertain the quality of the silver thin film, we illuminated it normally and measured the transmittance for two linear polarization states of the illuminating light: (i) p-polarization, when the electric field has a component parallel to the nanorods, and (ii) s-polarization, when the electric field is perpendicular to the nanorods, as shown in Fig. 1. Figure 2 contains the spectra of the two transmittances, $T_p$ and $T_s$, over the 300-to-850-nm wavelength range. The peaks at $\lambda$ = 324 nm and 321 nm, respectively, of $T_p$ and $T_s$ are in accord with measured and calculated spectra of absorbances of silver thin films.[15]

The silver thin film, when considered as a continuum at sufficiently large wavelengths, has to be orthorhombic. It should have linear dielectric and magnetic properties. In general, the effective relative permittivity tensor $\vec{\varepsilon}$ of the film then has three distinct eigenvalues, and so must the effective relative permeability tensor $\vec{\mu}$, but both will have the same set of three eigenvectors. When the film is illuminated normally, different combinations of the eigenvalues of these tensors appear for the two polarization states. We label the combinations $\varepsilon_p$ and $\mu_p$ for p-polarization, and $\varepsilon_s$ and $\mu_s$ for s-polarization.

For a specific linear polarization state, the refractive index

$n_v = \sqrt{\varepsilon_v \mu_v} = n'_v + in''_v$ and the relative intrinsic impedance $\eta_v = \sqrt{\mu_v/\varepsilon_v} = \eta'_v + i\eta''_v$ are complex-valued functions of the wavelength, where $i = \sqrt{-1}$ and $v = p, s$. Both $n_v$ and $\eta_v$ can be determined after measuring the reflection coefficient $r_v$ and the transmission coefficient $\tau_v$ of the silver thin film of thickness $d$ as follows:[16,17]

$$\eta_v = \pm\sqrt{\frac{(1+r_v)^2 - \tau_v^2}{(1-r_v)^2 - \tau_v^2}} \text{ with } \eta'_v > 0, \quad (1)$$

$$n_v = \frac{\lambda}{2\pi d}\cos^{-1}(\frac{1 - r_v^2 + \tau_v^2}{2\tau_v}). \quad (2)$$

The two reflection coefficients $r_{p,s}$ and the two transmission coefficients $\tau_{p,s}$ were measured at specific wavelengths using an ellipsometer[18] and a walk-off interferometer.[19] A diode laser operating at 532-nm, 639-nm or 690-nm wavelength was used as the source of monochromatic light. The ellipsometer used is a PSA (Polarizer-Sample-Analyzer) system built by J.A. Woollam Co. When normally incident light passes through the PSA system, the ellipsometric parameters can be measured by detecting the transmitted light as the analyzer rotates. The measured ellipsometric parameters yield the ratio $\tau_p/\tau_s$.

Walk-off interferometry was used to measure $\tau_s$ and both reflection coefficients. In this technique, the incident laser beam is separated into two beams—one s-polarized and the other p-polarized. One of the polarized beams is normally incident on the sample (silver thin film) and the other polarized beam is incident on the bare substrate; the two reflected and transmitted beams combine and produce interference. The polarization state of the combined beam yields the absolute phase of the reflection coefficient or the transmission coefficient associated with a specific polarization state. Finally, $\tau_p$ can be derived from the measured $\tau_s$ and the ratio $\tau_p/\tau_s$.

The values of the refractive indices $n_{p,s}$ and the relative intrinsic impedances $\eta_{p,s}$ derived from Eqs. (1) and (2) using the measured values of $r_{p,s}$ and $\tau_{p,s}$ at the three wavelengths are presented in Table 1. Since none of the six data sets satisfy either the condition $n_v = 1/\eta_v$ or the condition $n_v = \eta_v$, the silver thin film must have both dielectric and magnetic properties different from those of vacuum. The relative permittivities $\varepsilon_{p,s}$ and relative permeabilities $\mu_{p,s}$, calculated using the relationships $\varepsilon_v = n_v/\eta_v$ and $\mu_v = n_v\eta_v$, as listed in Table 1.

The imaginary parts of $\varepsilon_{p,s}$ and $\mu_{p,s}$ are positive in Table 1, which is appropriate for a passive material when an $\exp(-i\omega t)$ time-dependence is used with $\omega$ as the angular frequency and $t$ as time. As the real parts of both $\varepsilon_s$ and $\mu_s$ are also positive, the real part of $n_s > 0$ at all three wavelengths. However, while the real parts of $\mu_p$ are positive, the real parts of $\varepsilon_p$ are negative at all three wavelengths. The absolute real part and the imaginary part of $\varepsilon_p$ increase with wavelength, similar to that predicted by the plasmonic-type permittivity model for composites containing thin-wire metal inclusions.[20,21]

Consistently with the requirement that the imaginary part of $n_p > 0$ for attenuation in a passive medium,[22] we get the real part of the refractive index for p-polarization to be negative over a wide range of wavelengths in the visible regime: $n'_p = -0.705$ at $\lambda = 532$ nm, $n'_p = -0.476$ at $\lambda = 639$ nm, and $n'_p = -0.552$ at $\lambda = 690$ nm. Clearly, $n'_p < 0$ for light of green, yellow, orange, and red colors. This is our chief result.

The measured figure of merit $-n'_p/n''_p$ ranges between 0.3 and 0.65 over this range. This could be adjusted to more favorable values for device application by suitably changing the deposition angle and/or by filling in the void spaces in the thin film with a gain medium, both of which are topics for further research. Although our finding of a negative real part of a refractive index strictly holds for normal illumination, relying to the analytic continuation of reflection and transmission coefficients with respect to the angle of illumination, we expect negative refraction to be shown for p-polarized light for a restricted range of oblique-illumination conditions as well. Further characterization of the silver thin films is planned.

## 3. CONCLUDING REMARKS

We conclude that a widely used and a well-established thin-film technique called oblique angle deposition can be used to deposit thin films that can refract visible light negatively, and that this negative refraction can occur over a broad range of wavelengths. The OAD technique is particularly suitable for depositing multilayered stacks commonly employed as optical filters and mirrors.[4] Incorporation of layers of a gain medium to offset attenuation[23] is easily possible with the OAD technique.

**Acknowledgements**
YJJ, CWY and JTL thank the National Science Council of the Republic of China, Taiwan, for financially supporting this research under Contract No. NSC 96-2221-E-027-051-MY3. AL thanks the Charles Godfrey Binder Endowment at Penn State for partial support.

**Table 1. Effective refractive indices, relative intrinsic impedances, relative permittivities, and relative permeabilities for a silver thin film comprising parallel tilted nanorods.**

| $\lambda$ (nm) | $n_p$ | $\eta_p$ | $\varepsilon_p$ | $\mu_p$ |
|---|---|---|---|---|
| 532 | -0.705 + $i$ 1.091 | 1.657 - $i$ 1.731 | -0.533 + $i$ 0.102 | 0.721 + $i$ 3.030 |
| 639 | -0.476 + $i$ 1.586 | 0.330 - $i$ 0.719 | -2.073 + $i$ 0.290 | 0.983 + $i$ 0.865 |
| 690 | -0.552 + $i$ 1.755 | 0.271 - $i$ 0.550 | -2.967 + $i$ 0.456 | 0.816 + $i$ 0.778 |
| | $n_s$ | $\eta_s$ | $\varepsilon_s$ | $\mu_s$ |
| 532 | 0.386 + $i$ 0.516 | 0.917 - $i$ 0.356 | 0.176 + $i$ 0.632 | 0.538 + $i$ 0.336 |
| 639 | 0.520 + $i$ 0.457 | 0.984 - $i$ 0.154 | 0.445 + $i$ 0.534 | 0.582 + $i$ 0.370 |
| 690 | 0.404 + $i$ 0.450 | 0.933 - $i$ 0.236 | 0.293 + $i$ 0.556 | 0.483 + $i$ 0.325 |

**Figure captions**

Figure 1 (a) Schematic indicating the normal illumination by linearly polarized light of a silver thin film comprising parallel tilted nanorods. The electric field of the illuminating light either has a component parallel to the nanorods ($\vec{E}_p$).or is perpendicular to the nanorods ($\vec{E}_s$). (b) Two scanning-electron-microscopic (SEM) images of the silver thin film.

Figure 2 Measured spectra of the transmittances $T_{p,s} = |\tau_{p,s}|^2$ when the silver thin film is illuminated normally.

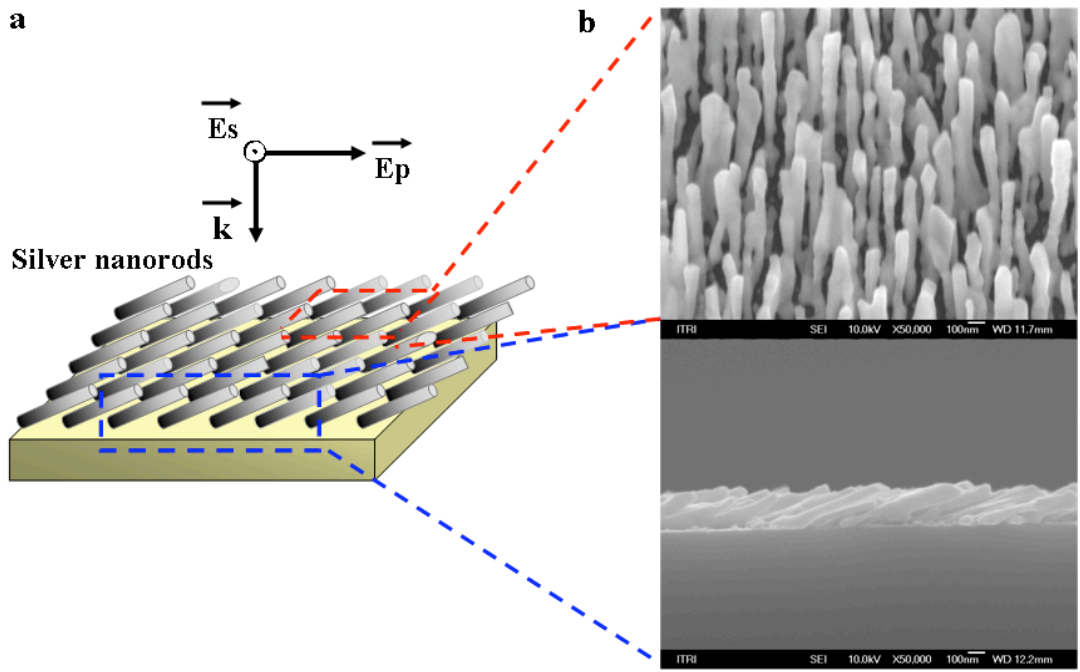

Figure 1.

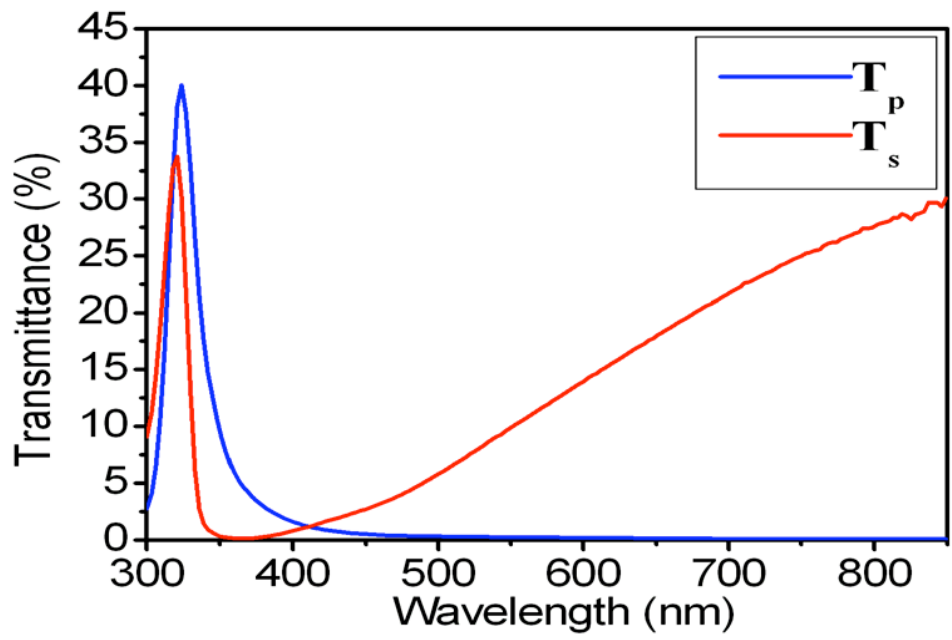

Figure 2.